\documentclass[12pt,a4paper]{article}
\begin{document}
{\Large On generation of the Bargmann-Moshinsky basis of $\mathrm{SU(3)}$ group}
\vspace{5mm}

\centerline{\large \it S. Vinitsky$^{~a,b,}$\footnote{e-mail: vinitsky@theor.jinr.ru}, \v C.~Burdik$^{~a,c}$, A.~Gusev$^{~a}$, A. Deveikis$^{~d}$,}  \centerline{\large \it A. G\'o\'zd\'z$^{~e}$, A.~P\c{e}drak$^{~f}$, P.M. Krassovitskiy$^{~a,g}$}

\vspace{5mm}
\noindent 
{\it$^{a}$Joint Institute for Nuclear Research, Dubna, Russia\\ 
 $^{b}$RUDN University, 6 Miklukho-Maklaya, 117198 Moscow, Russia\\
 $^{c}$Department of Mathematics
Faculty of Nuclear Sciences and Physical Engineering
Czech Technical University, Prague,
Czech Republic\\
 $^{d}$ Department of Applied Informatics, Vytautas Magnus University,
  Kaunas, Lithuania\\
 $^{e}$Institute  of Physics, Maria Curie{-}Sk{\l}odowska University, Lublin, Poland\\
 $^{f}$National Centre for Nuclear Research, Warsaw, Poland\\
 $^{g}$Institute of Nuclear Physics, Almaty, Kazakhstan}

\def\bxi{\mbox{\boldmath$\xi$}}
\def\bta{\mbox{\boldmath$\eta$}}
\newcommand{\Group}[1]{\textrm{#1}}

\newcommand{\MatElemSU}[3]{
\left.\left\langle \begin{array}{l} #1 \end{array} \right|
 #2
 \left| \begin{array}{l} #3 \end{array} \right\rangle\right.}

\newcommand{\RedMatElemSU}[3]{
\left.\left\langle \begin{array}{l}  #1 \end{array} \right|\right|
 #2
 \left|\left| \begin{array}{l} #3 \end{array} \right\rangle\right.}

\begin{abstract}
  An efficient procedure of orthonormalisation of the Bargmann--Moshinsky
  (BM) basis
  %corr26102018
  is examined using analytical formulas of the overlap integrals of
  the BM basis.  Calculations of components of the quadrupole operator between
  the both BM and the orthonormalised bases needed for construction of the
  nuclear models are tested.  The proposed procedure is also implemented as the
  Fortran program.\footnote{Submitted to: Journal of Physics: Conference Series }
\end{abstract}

%%%%%%%%%%%%%%%%%%%%%%%%%%%%%%%%%%%%%%%%%%%%%%%%%%%%%%%%%%%%%%%%%%%%%
\section{Introduction}

The formalism of $\mathrm{SU(3)}$ group provides a comprehensive theoretical
foundation for understanding this symmetry in nuclear structure
\cite{Cseh15,Dytrych16,Ell58,Gozdz18,Harvey}.  However, the construction of the
$\mathrm{SU(3)}$ bases usually can be performed analytically only for some
special cases.  In this respect, because of mathematical simplicity of its
definition, the Bargmann--Moshinsky (BM) basis \cite{BarMos61,MosPatShaWin75} is
especially convenient for calculation.  However, the necessity to introduce the
physically relevant angular momentum observable gives rise to the non-canonical
group reduction $\mathrm{SU(3)} \supset \mathrm{SO(3)} \supset
\mathrm{SO(2)}$. The BM vectors may be calculated from the simplest vectors
which correspond to the highest angular momentum projection $M=L$, i.e.  the
highest weight basis vectors with respect to the $\Group{SO(3)}$ {group that was
  proved in \cite{MosPatShaWin75}}. It should be stressed, that the analytical
and what is very important an { effective} algorithm for construction of this
basis is required for analysis of some quantum systems.

As an example one can consider the quadrupole vibrational and rotational motions
which are the most important low energy nuclear motions. The simplest
$\mathrm{SU}(3)$ model Hamiltonian consists of the quadrupole-quadrupole
interaction, the rotational term and potentially the other terms constructed
from generators of the partner groups $G=\mathrm{SU(3)} \times
\overline{\mathrm{SU(3)}}$, see \cite{Gozdz18} and references therein. A
possible Hamiltonian $H$ used in this schematic nuclear model can be written as:
\begin{eqnarray}
 H= \gamma C_2(\mathrm{SU(3)})  -\kappa Q \cdot Q + \beta L \cdot L
+H''(\bar{Q},\bar{L})   \nonumber \\
  =(\gamma-\kappa) C_2(\Group{SU(3)}) +
(3\kappa+\beta) L^2+H''(\bar{Q},\bar{L}),
\label{H_SU3}
\end{eqnarray}
where the second order Casimir operator $C_2(\mathrm{SU(3)})= Q \cdot Q +3 L
\cdot L$, $Q$ and $L$ are generators of $\mathrm{SU(3)}$, i.e. quadrupole and
  angular momentum, respectively. $\bar{Q}$ and $\bar{L}$ are generators of the
intrinsic group $\overline{\mathrm{SU(3)}}$.
Some examples of physically interesting forms of the interaction $H''$ can be
written as
\begin{eqnarray}
\label{TetrahedrTermsHam}
&& H_{3Q}=h_{3Q} \left( (\bar Q \otimes \bar Q)^3_{2}-(\bar Q \otimes \bar Q)^3_{-2} \right)
\, , \\
&& H_{3LQ}=h_{3LQ} \left( (\bar L \otimes \bar Q)^3_{2}-(\bar L \otimes \bar Q)^3_{-2} \right)
\, , \\
&& H_{4Q}= h_{4Q} \Big( \sqrt{\frac{14}{5}}(\bar Q \otimes \bar Q)^4_{0}
+ (\bar Q \otimes \bar Q)^4_{-4} + (\bar Q \otimes \bar Q)^4_{4} \Big) ,
\end{eqnarray}
where $(T_{\lambda'} \otimes T_{\lambda})^L_{M}$ denotes the tensor product of
two spherical tensors \cite{Varshalovitch}. For, example, these interaction
terms can simulate either the tetrahedral or octahedral nuclear symmetry now
widely considered in nuclear physics \cite{Dudek2002}. To find the corresponding
energies and quantum nuclear states one needs to solve the eigenvalue problem of
the Hamiltonian (\ref{H_SU3}).

To solve the eigenvalue problem for $H$ the appropriate basis constructed
according to the group chain $\Group{SU(3)} \supset \Group{SO(3)} \supset
\Group{SO(2)} $ is required. There were several attempts to construct such
bases. They were based on different group theoretical technics, for a short
review see the introduction in the paper \cite{Pan16,AliRayRos81}.  In all those cases one
obtains the non-orthogonal basis. This increases a complexity of calculations of
the reduced matrix elements of different operators, Clebsch-Gordan coefficients,
etc.
It requires an adaptation of the Gram-Schmidt like orthogonalization procedure
to be more effective in symbolic calculations.

We start from the BM states which are linearly independent but as in other
approaches not orthonormal. We developed an { effective} symbolic algorithm
{suitable} for implementation in computer algebra systems \cite{casc18}.  It is based on the
adapted Gram-Schmidt orthonormalization procedure but {using the overlap}
integrals calculated in an analytical form \cite{AliRayRos81}. { It provides the
  analytic construction of the desirable orthonormalized basis.}  {Our
  adaptation of Gram-Schmidt orthonormalization procedure consists in
  construction of recursive calculation of the required quantities and the
  normalization integrals. These calculations do not involve any square root
  operation. This distinct features of the proposed orthonormalization algorithm
  allows for a large scale symbolic calculations \cite{casc18}.}

Then one can calculate in { this} orthonormalized basis the zero component of
the quadrupole operator $Q_0$ in the analytical form using its simpler form
given in the non-canonical BM basis~\cite{Afanasjev,Raychev1981}. The other
components of the quadrupole operator $Q_{k}$ written in the analytical form can
be obtained by making use of the Wigner-Eckart theorem with conventional
$\mathrm{SO(3)}$ Clebsch-Gordan coefficients \cite{Varshalovitch}.  This is
required theorem for a final construction of the above Hamiltonian (\ref{H_SU3})
also in an analytical form.

Meanwhile, to organize a real large scale calculations of the required matrix
elements of the quadrupole operator with $\lambda,\mu>5$, where $(\lambda,\mu)$
are Elliot labels denoting the irreducible representations (irrep.) of
$\mathrm{SU(3)}$, one needs to have a quick algorithm implemented in Fortran
that reduces computer resources. The construction and testing of such algorithm
is given in the present paper.

The paper is organized as follows.  In Section 2, the \textit{procedure 1} for
calculation of overlap integrals of {BM} vectors is shown.  In Section 3, the
\textit{procedure 2} for orthonormalization of BM basis is given. In Section 4,
the \textit{procedure 3} of an action of the quadrupole operator $Q_0$ onto the
constructed basis is presented.  In Conclusions further applications of the
elaborated \textit{ procedures} are outlined.

%%%%%%%%%%%%%%%%%%%%%%%%%%%%%%%%%%%%%%%%%%%%%%%%%%%%%%%%%%%%%%%%%%%%%%%%
\section{Calculations of the overlap integrals of the BM basis }

The { effective} method for constructing a non-canonical BM basis with the
highest weight vectors of $\mathrm{SO(3)}$ irreducible representations
corresponding to the group chain $SU(3)\supset O(3)\supset O(2)$ with
commutation relations of the spherical tensors $L_\nu(\nu=\pm1,0)$,
$Q_\nu(\nu=\pm2,\pm1,0)$:
\begin{eqnarray}\label{comrelsu3}
[L_{\nu},L_{\nu'}]=-\sqrt2C_{1\nu1\nu'}^{1\nu+\nu'}L_{\nu+\nu'},~ \\ \nonumber
[L_\nu,Q_{\nu'}]=-\sqrt6C_{1\nu2\nu'}^{2\nu+\nu'}Q_{\nu+\nu'},~\\ \nonumber
[Q_\nu,Q_{\nu'}]=-3\sqrt{10}C_{2\nu2\nu'}^{1\nu+\nu'}L_{\nu+\nu'},~
\end{eqnarray}
and the Casimir operator
$$C_{2}(SU(3))= Q\cdot Q+3{  L}\cdot {  L}=4(\lambda^2+\mu^2+\lambda\mu+3\lambda +3\mu)$$ was described in
\cite{AliRayRos81} and implemented as a symbolic algorithm in \cite{casc18}.
   Let us introduce the notation for overlaps of the vectors of this basis:
\begin{equation}\label{BMstate}
 \biggl\langle u_\alpha\biggl| u_{\alpha'} \biggr\rangle=
\left\langle
\begin{array}{l}
(\lambda,\mu)_B\\
\alpha,L,M\end{array}
\right.
\left|
\begin{array}{l}
(\lambda,\mu)_B\\
\alpha',L,M\end{array}\right\rangle
\end{equation}
Here the quantum numbers $\lambda,\,\mu$ which label { the irreducible
  representations (irreps)}, $\lambda,\mu = 0,1,2,\dots$ and $\lambda > \mu$;
$L, M$ are the quantum numbers of angular momentum ${L}\cdot {L}$ and
its projection $L_0$ (in our case, $M=L$); $\alpha$ is the additional index that
is used for unambiguously distinguishing the equivalent $\mathrm{SO(3)}$ irreps
$(L)$ in a given $\mathrm{SU(3)}$ irrep $(\lambda,\,\mu)$.  The dimension of
an irrep of SU(3) for a given $\lambda,\mu$ can be calculated by using the following
formula:
 \begin{equation}
 \label{dimfa}
 D_{\lambda\mu}=\frac{1}{2}(\lambda+1)(\mu+1)(\lambda+\mu+2).
 \end{equation}
In order to perform classification of the BM states (\ref{BMstate}) one should
determine a set of allowed values of $\alpha$ and $L$.  It is well known that
the ranges of quantum numbers $\alpha$ and $L$ are determined by the values of
quantum numbers $\lambda$ and $\mu$.
However, the determination of the former quantities is rather cumbersome.

 {The easiest way} to get the allowed values
of $\alpha$ and $L$ is by using the following \textit{procedures}:

{\it Step 1.} Firstly we start with choosing some particular value of the quantum
number $\mu$. For the following consideration, it is convenient to introduce
auxiliary label $K$ \cite{Ell58} which varies within the range
\begin{equation}
\label{Kdef}
K=\mu,\mu-2,\mu-4, ..., 1 \mbox{ or } 0, \quad \mbox{ for } \lambda > \mu.
\end{equation}
The label $K$ is related to $\alpha$ by
\begin{equation}
\alpha=\frac{1}{2}(\mu-K).
\label{alphadef}
\end{equation}
So, for every fixed $\mu$, the set of possible values of $K$ can be obtained
directly from Ref. (\ref{Kdef}). Now, the set of allowed values
of $\alpha$ may be determined from these $K$ values using  relation
(\ref{alphadef}).

{\it Step 2.} In the case $K = 0$, that may occur only for even values of $\mu$, the
allowed values of $L$ are determined by the label $\lambda$:
\begin{equation}
\label{LK0def}
L=\lambda,\lambda-2,\lambda-4, ..., 1 \mbox{ or } 0.
%%\nonumber
\end{equation}
{\it Step 3.} In the case $K \ne 0$ the $L_\mathrm{min}=K$. Since for every
particular $\mu$, there is a number of possible $K$ numbers, according to
(\ref{Kdef}) there exists a number of the corresponding numbers  $\alpha$. It
means that for every particular $\mu$, there will be a number of pairs
$(\alpha, L_\mathrm{min})$.
The maximal value of $L$ is defined by the expression
$L_\mathrm{max}=\mu-2\alpha+\lambda-\beta$, where
\begin{equation}
\label{betadef}
\beta=\left\{
\begin{array}{l}
	0, \quad \lambda + \mu -L \mbox{ even},\\
	1, \quad \lambda + \mu -L \mbox{ odd}.
\end{array}
\right.
%\nonumber
\end{equation}
To determine $L_\mathrm{max}$ it is convenient to consider two alternatives:
$\lambda -L \mbox{ is even}$ and $\lambda -L \mbox{ is odd}$. In both cases, the
label $\beta$ is defined by the given value $\mu$ value. The number
$L_\mathrm{max}$ is determined in a similar manner.  An illustrative example for
calculation of allowed values of $\alpha$ and $L$ is presented in Table 1 of ref
\cite{casc18}.
It should be noted that the set of allowed values of $L$ for overlap integrals
is given by intersection of these sets for the corresponding $<bra|$ and
$|ket>$ vectors.

The highest-weight vector of the BM basis of the SO(3) multiplets for all kinds
of irreducible representation of SU(3) can be written in the form
\cite{AliRayRos81}
\begin{eqnarray}
\left|\begin{array}{c} (\lambda,\mu)_B \\
\alpha~L~L\end{array}\right\rangle\equiv
|\beta n_0 n_1 n_2 \alpha \rangle
=w^\beta (\xi_1)^{n_0}(x_{10})^{n_1}(\bxi^2)^{n_2}A^\alpha|0 \rangle,
\label{bm1}
\end{eqnarray}
that differ from the states Eq.  (3.8) given in \cite{MosPatShaWin75} in the
definition of the number $\alpha$ and coincide up to a phase factor
$(-1)^{\alpha}$.  Here the set of numbers $n_0$, $n_1$, $n_2$ is given in terms
of the above defined $\lambda$, $\mu$, $\alpha$, $L$ and $\beta$
 \begin{eqnarray}
n_0=L-\mu+2\alpha,\quad n_1=\mu-2\alpha-\beta,\quad n_2=
(\lambda+\mu-L-2\alpha-\beta)/2.
\end{eqnarray}
The corresponding  operators are determined by the following relations:
\begin{eqnarray}\label{bm1a}
&&x_{10}=\xi_1\eta_0-\xi_0\eta_1,\quad
x_{1-1}=\xi_1\eta_{-1}-\xi_{-1}\eta_1,\quad
x_{0-1}=\xi_0\eta_{-1}-\xi_{-1}\eta_0,\\ &&
w=\xi_1x_{1-1}-\xi_0x_{10}=
\xi_1^2\eta_{-1}-\xi_{-1}\xi_1\eta_{1}-\xi_{0}\xi_1\eta_{0}+\xi_0^2\eta_1,\quad
\bxi^2=\xi_0^2-2\xi_{-1}\xi_1, \nonumber \\  && A=(2x_{10}x_{0-1}-x_{1-1}^2) \nonumber\\  &&
=
 2\xi_0\xi_1\eta_0\eta_{-1}-2\xi_{-1}\xi_1\eta_0^2-2\xi_0^2\eta_{-1}\eta_1
+2\xi_0\xi_{-1}\eta_0\eta_1
 {-}\xi_1^2\eta_{-1}^2
 {+}2\xi_{-1}\xi_1\eta_{-1}\eta_1
 {-}\xi_{-1}^2\eta_1^2, \nonumber
\end{eqnarray}
via two SO(3) spherical vectors belong to two independent SU(3) representations
\cite{Varshalovitch}
\begin{eqnarray}\label{15}
\xi_\pm=\mp\frac1{\sqrt2}(\xi_x\pm\imath\xi_y), \quad\xi_0=\xi_z,
\end{eqnarray}
which we consider as the vector-boson creation operators $\xi_m$ and $\eta_m$
with $(L,M)=(1,M=1,0,-1)$.  Then the states (\ref{bm1}) are polynomials
constructed from these operators which act on the vacuum state denoted by
$|0\rangle$.
The pairs of creation $\xi_m$ and $\eta_m$, and annihilation
$\xi_{m}^{+}$ and $\eta_{m}^{+}$ vector-boson operators are defined by relations
\begin{eqnarray}\label{16}
\xi_{m}^{+}|0 \rangle=\eta_{m}^{+}|0 \rangle=0, \quad
[\xi_{m}^{+},\xi_{n}]=[\eta_{m}^{+},\eta_{n}]=(-1)^{m}\delta_{-m,n}.
\end{eqnarray}
With the help of $\bxi$ and $\bta$ one can construct the irreducible tensor
operators
$$F_M^L=\sum_{\mu\nu}C_{1\mu1\nu}^{LM} (\xi_\mu\xi_\nu^++\eta_\mu\eta_\nu^+),
$$
where $C_{1\mu1\nu}^{LM}$ are Clebsch--Gordan coefficients
\cite{Varshalovitch}. The vectors $\bxi^+$ and $\bta^+$ can be chosen in the
form
$$\xi_\nu^+=(-1)^{\nu}\partial/\partial\xi_{-\nu},
\quad \eta_\nu^+=(-1)^{\nu}\partial/\partial\eta_{-\nu},$$
i.e. the vectors $\bxi$, $\bta$ and $\bxi^+$, $\bta^+$ can be considered as
creation and annihilation operators of two distinct kinds of vector-boson in
Fock representation.

The tensor operators satisfy the following commutation relations
$$
[F_{M_1}^{L_1},F_{M_2}^{L_2}]=\sqrt{(2L_1+1)(2L_2+1)}\sum_L((-1)^{L_1+L_2}-(-1)^L)C_{L_1M_1L_2M_2}^{LM_1+M_2}
\left\{\begin{array}{ccc}L_1&L_2&1\\1&1&1\end{array}\right\} F_{M_1+M_2}^L,$$
where $\{ ...\}$ is the 6j--Wigner symbol \cite{Varshalovitch}.  If we introduce
$L_m=-\sqrt2F_m^1$ and $Q_k=-\sqrt6F_k^2$, we can see that operators $L_m$
$(m=0,\pm1)$ and $Q_k$ $(k=0,\pm1,\pm2)$ satisfy the standard commutation
relations of SU(3)group (\ref{comrelsu3}).
It is evident that the operators $L_m$ $(m=0,\pm1)$ define the algebra of
angular momentum SO(3) and the operators $Q_k$ $(k=0,\pm1,\pm2)$ extend this
algebra to SU(3) algebra.

Using the above definitions (\ref{15}) and (\ref{16}), we determine the standard boson basis
\begin{eqnarray}
|k_1,k_2,k_3,k_4,k_5,k_6\rangle=(k_1!k_2!k_3!k_4!k_5!k_6!)^{-1/2}
(\xi_{-1})^{k_1}(\xi_{0})^{k_2}(\xi_{1})^{k_3}
(\eta_{-1})^{k_4}(\eta_{0})^{k_5}(\eta_{1})^{k_6}|0 \rangle, \label{bm2}
\end{eqnarray}
which is orthonormal
\begin{eqnarray}
\langle k_1',k_2',k_3',k_4',k_5',k_6'|k_1,k_2,k_3,k_4,k_5,k_6\rangle=
\delta_{k_1'k_1}\delta_{k_2'k_2}\delta_{k_3'k_3}\delta_{k_4'k_4}\delta_{k_5'k_5}\delta_{k_6'k_6}.
\end{eqnarray}

Then we expand the vectors (\ref{bm1}) in the terms of basis (\ref{bm2}). As the first
step we apply the multipliers of (\ref{bm1}) in the variables $\xi_{-1}$,
$\xi_{0}$, $\xi_{1}$, $\eta_{-1}$, $\eta_{0}$, $\eta_{1}$.  In fact we use only
operator expansion. Because $\beta=0$ or $\beta=1$, we write
\begin{eqnarray}\label{wbeta}
  w^\beta=
\sum_{\nu}b_\nu^\beta (\xi_{-1})^{\nu_1}(\xi_{0})^{\nu_2}(\xi_{1})^{\nu_3}
  (\eta_{-1})^{\nu_4}(\eta_{0})^{\nu_5}(\eta_{1})^{\nu_6},\quad
\nu\equiv (\nu_1,\nu_2,\nu_3,\nu_4,\nu_5,\nu_6).
\label{bm3}
\end{eqnarray}
By comparing (\ref{bm1a}) and (\ref{bm3}), we obtain that the sum in (\ref{bm3})
contains one term for $\beta=0$ and four terms for $\beta=1$:
$$ b_{(0, 0, 0, 0, 0, 0)}^0{=}1,\, b_{(0, 0, 2, 1, 0, 0)}^1{=}1,\,
b_{(1, 0, 1, 0, 0, 1)}^1{=}-1,\, b_{(0, 1, 1, 0, 1,0)}^1
{=} -1,\, b_{(0, 2, 0, 0, 0, 1)}^1{=}1.
$$
From (\ref{bm1a}) using the multinomial theorem, we also calculate the needed
powers of the operators
\begin{eqnarray}\label{x10}
(x_{10})^{n_1}%=(\xi_1\eta_0-\xi_0\eta_1)^n_1
=\sum_{k_1=0}^{n_1}\left(\begin{array}{c} n_1 \\ k_1\end{array}\right) (-1)^{k_1}\xi_0^{k_1}\xi_1^{n_1-k_1}\eta_0^{n_1-k_1}\eta_1^{k_1},\\ \nonumber
(\bxi^2)^{n_2}%=(\xi_0^2-2\xi_{-1}\xi_1)^n_n
=\sum_{k_2=0}^{n_2}\left(\begin{array}{c} n_2 \\ k_2\end{array}\right) (-1)^{k_2}2^{k_2}\xi_{-1}^{k_2}\xi_0^{2(n_2-k_2)}\xi_1^{k_2}\eta_0^{n_1-k_1},\\ \nonumber
A^\alpha=\sum_{s\in\Omega_{s,\alpha}}
\left(\begin{array}{c} \alpha \\ s_1s_2s_3s_4s_5s_6s_7 \end{array}\right)
(-1)^{s_1+s_3+s_5+s_7}
\xi_{-1}^{s_2+s_4+s_6+2s_7}
\xi_{0}^{s_1+2s_3+s_4}
\\\nonumber
\times
\xi_{1}^{s_1+s_2+2s_5+s_6}
\eta_{-1}^{s_1+s_3+2s_5+s_6}
\eta_{0}^{s_1+2s_2+s_4}
\eta_{1}^{s_3+s_4+s_6+2s_7},
%\\\nonumber
% \Omega_{s,\alpha}=\{s_1,...,s_7|s_1\geq0,...,s_7\geq0,s_1+...+s_7=\alpha\}.
\end{eqnarray}
where a set of indices runs within the range
$\Omega_{s,\alpha}=\{s_1,...,s_7|s_1\geq0,...,s_7\geq0,s_1+...+s_7=\alpha\}$.

So, we have the highest-weight vector of the BM basis
\begin{eqnarray} \label{bm4}
|\beta n_0 n_1 n_2 \alpha \rangle
=\sum_{\nu}\sum_{k_1=0}^{n_1}\sum_{k_2=0}^{n_2}\sum_{s\in\Omega_{s,\alpha}}
b_\nu^\beta (-1)^{k_1+k_2+s_1+s_3+s_5+s_7} 2^{k_2+s_1+s_2+s_3+s_4+s_6}
\\ \nonumber
\times\left(\begin{array}{c} n_1 \\
k_1\end{array}\right)\left(\begin{array}{c}  n_2 \\
k_2\end{array}\right)\left(\begin{array}{c} \alpha \\
s_1s_2s_3s_4s_5s_6s_7 \end{array}\right)
(\xi_{-1})^{\gamma_1}(\xi_{0})^{\gamma_2}(\xi_{1})^{\gamma_3}
(\eta_{-1})^{\gamma_4}(\eta_{0})^{\gamma_5}(\eta_{1})^{\gamma_6}|0\rangle,
\end{eqnarray}
where the set multi-indices $\gamma_1$,...,$\gamma_6$ is determined by the
relations
\begin{eqnarray}
\gamma_1={\nu_1+k_2+s_2+s_4+s_6+2s_7},\quad
\gamma_2={\nu_2+k_1+2(n_2-k_2)+s_1+2s_3+s_4},\\\nonumber
\gamma_3={\nu_3+n_0+n_1-k_1+k_2+s_1+s_2+2s_5+s_6},\quad
\gamma_4={\nu_4+s_1+s_3+2s_5+s_6},\\ \nonumber
\gamma_5={\nu_5+n_1-k_1+s_1+2s_2+s_4},\quad
\gamma_6={\nu_6+k_1+s_3+s_4+s_6+2s_7}.
\end{eqnarray}
From (\ref{bm4}) in the boson representation (\ref{bm2}) we obtain the required
BM states
\begin{eqnarray} \label{bm4a}
|\beta n_0 n_1 n_2 \alpha \rangle
=\sum_{\nu}\sum_{k_1=0}^{n_1}\sum_{k_2=0}^{n_2}\sum_{s\in\Omega_{s,\alpha}}
B(\beta,n_0,n_1,n_2,\alpha,\nu,k_1,k_2,s)
|\gamma_1,\gamma_2,\gamma_3,\gamma_4,\gamma_5,\gamma_6\rangle,
\end{eqnarray}
where the coefficients $B(\beta,n_0,n_1,n_2,\alpha,\nu,k_1,k_2,s)$ have the
following form
\begin{eqnarray}\nonumber
B(\beta,n_0,n_1,n_2,\alpha,\nu,k_1,k_2,s)=
b_\nu^\beta (-1)^{k_1+k_2+s_1+s_3+s_5+s_7} 2^{k_2+s_1+s_2+s_3+s_4+s_6}
\\ \nonumber
\times\left(\begin{array}{c} n_1 \\
k_1\end{array}\right)\left(\begin{array}{c} n_2 \\
k_2\end{array}\right)\left(\begin{array}{c} \alpha \\
s_1s_2s_3s_4s_5s_6s_7 \end{array}\right)
\sqrt{\gamma_1!\gamma_2!\gamma_3!\gamma_4!\gamma_5!\gamma_6!}.
\end{eqnarray}
{\it Step 4.}  The overlap integrals are determined by the relation
\begin{eqnarray} \label{bm5}
\langle\beta' n_0' n_1' n_2' \alpha'|\beta n_0 n_1 n_2 \alpha \rangle
=\delta_{\beta'\beta}\sum_{\nu',k_1',k_2',s'}\sum_{\nu,k_1,k_2,s}
B(\beta',n_0',n_1',n_2',\alpha',\nu',k_1',k_2',s')\\ \nonumber
\times B(\beta,n_0,n_1,n_2,\alpha,\nu,k_1,k_2,s)
\delta_{\gamma_1\gamma_1'}\delta_{\gamma_2\gamma_2'}\delta_{\gamma_3\gamma_3'}
\delta_{\gamma_4\gamma_4'}\delta_{\gamma_5\gamma_5'}\delta_{\gamma_6\gamma_6'},
\end{eqnarray}
where domains of summation are determined by the  definitions (\ref{wbeta})
and (\ref{x10}).

{\it Output of Step 4.} The overlap integral (\ref{bm5}) has been calculated with algorithm
implemented in Fortran. The obtained results for $\lambda=0,...,10$ and
$\mu=1,...,10$ for $\lambda\geq\mu$ and corresponding sets of values $\alpha$
and $L$ (for example, see Table 1 of ref. \cite{casc18}) coincide up to 10 digits
with results of calculations obtained with help of symbolic algorithm
\cite{casc18} implemented in Wolfram Mathematica.

%%%%%%%%%%%%%%%%%%%%%%%%%%%%%%%%%%%%%%%%%%%%%%%%%%%%%%%%%%%%%%%%%%%%%%%%%%%
\section{Orthogonalisation of the BM basis}

Let us construct the orthonormal basis in the space spanned by the non-canonical
BM vectors (\ref{BMstate}), $(M=L)$. For this purpose, we propose a bit more
efficient form of the Gram--Schmidt orthonormalisation procedure
\begin{equation}
|z_i\rangle\equiv\left|\begin{array}{l}
(\lambda,\mu)\\
f_i,L,L\end{array}\right\rangle=
\sum_{\alpha=0}^{\alpha_{\mathrm{max}}}A_{i,\alpha}^{(\lambda,\mu)}(L)
\left|\begin{array}{l}
(\lambda,\mu)_B\\
\alpha,L,L\end{array}\right\rangle\equiv
\sum_{\alpha=0}^{\alpha_{\mathrm{max}}}A_{i,\alpha}^{(\lambda,\mu)}(L)|u_\alpha\rangle.
\label{ortonormExpan}
\end{equation}
Here multiplicity index $i$ is introduced to distinguish the orthonormalized
states. The symbols $A_{i,\alpha}^{(\lambda,\mu)}(L)$ denotes the matrix elements of the
%cor 20 10 2018
upper
triangular matrix of the BM basis orthonormalization coefficients. These coefficients
{fulfill} the following condition
\begin{equation}
\label{Ais0def}
A_{i,\alpha}^{(\lambda,\mu)}(L)= 0, \quad \mbox{ if } i > \alpha.
\end{equation}
Because the BM vectors (\ref{BMstate}) are linearly independent, one can require
the orthonormalization properties for the vectors (\ref{ortonormExpan})
\begin{equation}
\left\langle
\begin{array}{l}
(\lambda,\mu)\\
f_i,L,L\end{array}
\right.
\left|
\begin{array}{l}
(\lambda,\mu)\\
f_k,L,L\end{array}\right\rangle=
\delta_{ik}.
\label{OverlapMat}
\end{equation}

{\it Step 5.} {Gramian} of a set of BM eigenvectors
%begin cor 20 10 2018
$u_{\alpha_{\max}},...,u_0$ from r.h.s. of
(\ref{ortonormExpan}) in notations (\ref{BMstate})
\begin{eqnarray*}
G(u_{\alpha_{\max}},...,u_0)=
\left|
\begin{array}{cccc}
\langle u_{\alpha_{\max}}|u_{\alpha_{\max}}\rangle &\ldots
%\langle u_{\alpha_{\max}-1}|u_{\alpha_{\max}}\rangle &\ldots
&\langle u_{\alpha_{\max}}|u_1\rangle&\langle u_{\alpha_{\max}}|u_0\rangle\\
%\langle u_{\alpha_{\max}-1}|u_{\alpha_{\max}}\rangle &\ldots
\langle u_{\alpha_{\max}-1}|u_{\alpha_{\max}}\rangle &\ldots
&\langle u_{\alpha_{\max}-1}|u_1\rangle&\langle u_{\alpha_{\max}-1}|u_0\rangle\\
\vdots
&\ddots&\vdots&\vdots\\

%\langle u_{\alpha_{\max}}|u_1\rangle&\ldots&\langle u_1|u_1\rangle
\langle u_1|u_{\alpha_{\max}}\rangle&\ldots&\langle u_1|u_1\rangle
&\langle u_1|u_0\rangle\\
\langle u_0|u_{\alpha_{\max}}\rangle&\ldots&\langle u_0|u_1\rangle
&\langle u_0|u_0\rangle
\end{array}
\right|.
\end{eqnarray*}
{Set of orthogonal (not orthonormal) vectors are calculated in the following
  way}  \cite{Gantmacher}:
\begin{eqnarray*}
|y_{\alpha_{\max}}\rangle=|u_{\alpha_{\max}}\rangle,
\qquad \bar A_{\alpha_{\max},\alpha_{\max}}^{(\lambda,\mu)}(L)=1,
\\
\qquad \bar A_{\alpha_{\max},s}^{(\lambda,\mu)}(L)=0,\qquad s=0,...,\alpha_{\max}-1,
\\
\, |y_{\alpha_{\max}-1}\rangle =\left|
\begin{array}{cc}
\langle u_{\alpha_{\max}}|u_{\alpha_{\max}}\rangle&|u_{\alpha_{\max}}\rangle\\
\langle u_{\alpha_{\max}-1}|u_{\alpha_{\max}}\rangle&|u_{\alpha_{\max}-1}\rangle
\end{array}
\right|,
\quad \\
\bar A_{\alpha_{\max}-1,\alpha_{\max}-1}^{(\lambda,\mu)}(L)=\langle u_{\alpha_{\max}}|u_{\alpha_{\max}}\rangle,
\\
\bar A_{\alpha_{\max}-1,\alpha_{\max}}^{(\lambda,\mu)}(L)=-\langle u_{\alpha_{\max}-1}|u_{\alpha_{\max}}\rangle,
\\
\qquad \bar A_{\alpha_{\max}-1,s}^{(\lambda,\mu)}(L)=0,\qquad s=0,...,\alpha_{\max}-2,\\
...,
\\
|y_{\alpha_{\max}-t}\rangle=
\left|
\begin{array}{cccc}
\langle u_{{\alpha_{\max}}}|u_{\alpha_{\max}}\rangle
%&\langle u_{{\alpha_{\max}}}|u_{\alpha_{\max}-1}\rangle
&\ldots
&\langle u_{{\alpha_{\max}}}|u_{\alpha_{\max}-t+1}\rangle
&|u_{{\alpha_{\max}}}\rangle
\\
\langle u_{{\alpha_{\max}}-1}|u_{\alpha_{\max}}\rangle
%&\langle u_{{\alpha_{\max}}-1}|u_{\alpha_{\max}-1}\rangle
&\ldots
&\langle u_{{\alpha_{\max}}-1}|u_{\alpha_{\max}-t+1}\rangle
&|u_{{\alpha_{\max}-1}}\rangle
\\
 \vdots&\ddots&\vdots&\vdots
\\
\langle u_{\alpha_{\max}-t+1}|u_{\alpha_{\max}}\rangle
%&\langle u_{\alpha_{\max}-t+1}|u_{\alpha_{\max}-1}\rangle
&\ldots
&\langle u_{\alpha_{\max}-t+1}|u_{\alpha_{\max}-t+1}\rangle
&|u_{\alpha_{\max}-t+1}\rangle
\\
\langle u_{\alpha_{\max}-t}|u_{\alpha_{\max}}\rangle
%&\langle u_{\alpha_{\max}-t}|u_{\alpha_{\max}-1}\rangle
&\ldots
&\langle u_{\alpha_{\max}-t}|u_{\alpha_{\max}-t+1}\rangle
&|u_{\alpha_{\max}-t}\rangle
\end{array}
\right|,
\\
\\
\bar A_{\alpha_{\max}-t,\alpha_{\max}-t'}^{(\lambda,\mu)}(L)=
(-1)^{t+t'+1}\left|
\begin{array}{ccc}
\langle u_{{\alpha_{\max}}}|u_{\alpha_{\max}}\rangle
%&\langle u_{{\alpha_{\max}}}|u_{\alpha_{\max}-1}\rangle
&\ldots
&\langle u_{{\alpha_{\max}}}|u_{\alpha_{\max}-t+1}\rangle
\\
\langle u_{{\alpha_{\max}}-1}|u_{\alpha_{\max}}\rangle
%&\langle u_{{\alpha_{\max}}-1}|u_{\alpha_{\max}-1}\rangle
&\ldots
&\langle u_{{\alpha_{\max}}-1}|u_{\alpha_{\max}-t+1}\rangle
\\
 \vdots&\ddots&\vdots
\\
\langle u_{\alpha_{\max}-t'-1}|u_{\alpha_{\max}}\rangle
%&\langle u_{\alpha_{\max}-t'-1}|u_{\alpha_{\max}-1}\rangle
&\ldots
&\langle u_{\alpha_{\max}-t'-1}|u_{\alpha_{\max}-t+1}\rangle
%\\
%\vdots&\vdots&\ddots&\vdots&\vdots
\\
\langle u_{\alpha_{\max}-t'+1}|u_{\alpha_{\max}}\rangle
%&\langle u_{\alpha_{\max}-t'+1}|u_{\alpha_{\max}-1}\rangle
&\ldots
&\langle u_{\alpha_{\max}-t'+1}|u_{\alpha_{\max}-t+1}\rangle
\\
 \vdots&\ddots&\vdots
\\
\langle u_{\alpha_{\max}-t+1}|u_{\alpha_{\max}}\rangle
%&\langle u_{\alpha_{\max}-t+1}|u_{\alpha_{\max}-1}\rangle
&\ldots
&\langle u_{\alpha_{\max}-t+1}|u_{\alpha_{\max}-t+1}\rangle
\\
\langle u_{\alpha_{\max}-t}|u_{\alpha_{\max}}\rangle
%&\langle u_{\alpha_{\max}-t}|u_{\alpha_{\max}-1}\rangle
&\ldots
&\langle u_{\alpha_{\max}-t}|u_{\alpha_{\max}-t+1}\rangle
\end{array}
\right|,
\\
t'=0,...,t,
\\
\qquad \bar A_{\alpha_{\max}-t,s}^{(\lambda,\mu)}(L)=0,\qquad s=0,...,\alpha_{\max}-t-1,
\end{eqnarray*}
where $t=0,1,...,\alpha_{\max}$. In result we have a  set of the orthonormal vectors
and coefficients:
\begin{eqnarray*}
|z_i\rangle=
\frac{|y_i\rangle}{\sqrt{\langle y_i|y_i\rangle}}=
\frac{|y_i\rangle}{\sqrt{G_{i+1}G_i}},\quad
\langle y_i|y_i\rangle=G_{i+1}G_i,\quad
i=\alpha_{\max},...,0,\quad G_{\alpha_{\max}+1}=1,\\
A_{\alpha_{\max}-t,s'}^{(\lambda,\mu)}(L)=\frac{\bar A_{\alpha_{\max}-t,s'}^{(\lambda,\mu)}(L)}
{\sqrt{
G_{\alpha_{\max}-t+1}
G_{\alpha_{\max}-t'}
}}.
\end{eqnarray*}
%end cor 20 10 2018

{\it Output of Step 5.} The required set
%cor 22 10 2018
of the coefficients
 $A_{i,\alpha}^{(\lambda,\mu)}(L)$
 are the components of the orthonormal vector
$|z_i\rangle$,
at $i=\alpha_{\max},
\alpha_{\max}-1,...,0$ has been calculated with the above algorithm implemented in
Fortran.
Note, the orthonormalization procedure is performed
in the reverse order with respect to the one adopted in \cite{Gantmacher},
that allows us to obtain the same orthonormal basis as in the papers \cite{AliRayRos81,casc18}.
The obtained results for $\lambda,\mu=0,...,10$, where
$\lambda\geq\mu$  coincide up to 10 digits with results of calculations
obtained by the symbolic algorithm \cite{casc18} implemented in Wolfram
Mathematica.

\section{Action of the the quadrupole operator onto the orthonormal  basis}

{\it Step 6.} Following the paper \cite{Raychev1981}, we determine the action of the
zero component $Q_0$ of the second order generator of $\mathrm{SU(3)}$ group onto the
BM basis vectors
\begin{equation}
Q_0\left|\begin{array}{l}
(\lambda,\mu)_B\\
\alpha,L,L\end{array}\right\rangle=
\sum_{\stackrel{\scriptstyle k=0,1,2}{\scriptstyle s=0,\pm1}}a_{s}^{(k)}
\left|\begin{array}{l}
(\lambda,\mu)_B\\
\alpha+s,L+k,L\end{array}\right\rangle \, .
%\nonumber
\label{Q0-ask}
\end{equation}
The coefficients $a_{s}^{(k)}$ can be calculated as in \cite{Afanasjev} and
 in \cite{Raychev1981}. To calculate action of  $Q_0$ onto the orthogonal
BM basis vectors (\ref{ortonormExpan}), we determine the  inverse transformation
$\tilde{A}_{i,\alpha}^{(\lambda,\mu)}(L)$ taken from the formula
(\ref{ortonormExpan})
\begin{equation}
\label{BMOrthogTrans}
\left|\begin{array}{l}
(\lambda,\mu)_B\\
\alpha,L,L\end{array}\right\rangle=
\sum_{i=0}^{\alpha} \tilde{A}_{i,\alpha}^{(\lambda,\mu)}(L)
 \left|\begin{array}{l}
(\lambda,\mu) \\
f_i,L,L\end{array}\right\rangle,
\end{equation}
where the following relations take place
\begin{equation}
\label{CoeffAOrthog}
\sum_i \tilde{A}_{i,\alpha'}^{(\lambda,\mu)}(L)
A_{i,\alpha}^{(\lambda,\mu)}(L)= \delta_{\alpha',\alpha}
\quad \mathrm{and} \quad
\sum_\alpha \tilde{A}_{i',\alpha}^{(\lambda,\mu)}(L)
A_{i,\alpha}^{(\lambda,\mu)}(L)= \delta_{i',i}.
\end{equation}
Using (\ref{Q0-ask}), (\ref{BMOrthogTrans}), and (\ref{CoeffAOrthog}), we
obtain the action of the zero component $Q_0$ of the quadrupole operator onto the
orthogonal BM basis vectors as
\begin{equation}
Q_0\left|\begin{array}{l}
(\lambda,\mu)\\
f_i,L,L\end{array}\right\rangle=
\sum_{\stackrel{\scriptstyle j=0,...,\alpha_{\max}}{\scriptstyle k=0,1,2}}q_{i,j,k}^{(\lambda,\mu)}(L)
\left|\begin{array}{l}
(\lambda,\mu)\\
f_j,L+k,L\end{array}\right\rangle,
\label{Q00-ask}
\end{equation}
where the coefficients $q_{i,j,k}^{(\lambda,\mu)}(L)$ are calculated by the formula
\begin{equation}\label{qijk}
q_{i,j,k}^{(\lambda,\mu)}(L)=
\sum_{\stackrel{\scriptstyle \alpha=0,...,\alpha_{\max}}{\scriptstyle s=0,\pm1}}
{A}_{i,\alpha}^{(\lambda,\mu)}(L)
a_s^{(k)}
{\tilde{A}}_{j,(\alpha+s)}^{(\lambda,\mu)}(L+k),
\end{equation}
and ${\tilde{A}}_{i,\alpha}^{(\lambda,\mu)}(L)$ are elements of the inverse and
{ the transpose of } the matrix $A^{(\lambda,\mu)}$
\begin{equation}
{\tilde{A}}_{i,\alpha}^{(\lambda,\mu)}(L)=
{(A^{-1})}_{\alpha,i}^{(\lambda,\mu)}(L).
\end{equation}

%%%%%%%%%%%%%%%%%%%%%%%%%%%%%%%%%%%%%%%%%%%%%%%%%%%%%%%%%%%%%%%%%%%%%%%%
\subsection{ Calculations of the action of the quadrupole operator zero
  component onto the BM basis}

Let us calculate action of the operator $Q_0$ according Eq. (\ref{Q0-ask})
\begin{eqnarray} &&
\!\!\!\!Q_0\left|\begin{array}{c} (\lambda,\mu)_B \\ \alpha~L~L\end{array}\right\rangle
{=}\sum_{s=0,\pm1}\left\{
\tilde a_s^{(0)}\left|\begin{array}{c} (\lambda,\mu)_B \\ \alpha{+}s~L~L\end{array}\right\rangle
%\right.\\ \left.
{+}\tilde a_s^{(1)}\left|\begin{array}{c} (\lambda,\mu)_B \\ \alpha{+}s~L{+}1~L\end{array}\right\rangle
%\right.\\ \left.
{+}\tilde a_s^{(2)}\left|\begin{array}{c} (\lambda,\mu)_B \\ \alpha{+}s~L{+}2~L\end{array}\right\rangle
\right\}
\label{bm2n}\\&&
{=}\sum_{s=0,\pm1}\left\{
\tilde a_s^{(0)}\left|\begin{array}{c} (\lambda,\mu)_B \\ \alpha{+}s~L~L\end{array}\right\rangle
%\right.\\ \left.
{+}\tilde a_s^{(1)}L_{-1}\left|\begin{array}{c} (\lambda,\mu)_B \\ \alpha{+}s~L{+}1~L{+}1\end{array}\right\rangle
%\nonumber\right.\\&& \qquad\qquad\qquad \left.
{+}\tilde a_s^{(2)}(L_{-1})^2\left|\begin{array}{c} (\lambda,\mu)_B \\ \alpha{+}s~L{+}2~L{+}2\end{array}\right\rangle
\right\},\nonumber
\end{eqnarray}
where the coefficients $\tilde a_s^{(k)}\equiv \tilde a_s^{(k)}(L)$ are
related to $a_s^{(k)}\equiv a_s^{(k)}(L)$ by the factors:
\begin{eqnarray}\label{bm7n}
\tilde a_s^{(2)}=a_s^{(2)} \sqrt{L+2}\sqrt{2L+3},\qquad \tilde a_s^{(1)}=a_s^{(1)} \sqrt{L+1},\qquad \tilde a_s^{(0)}=a_s^{(0)}.
\end{eqnarray}
We rewrite the BM basis vectors (\ref{bm1}) in the form
\begin{eqnarray}
\left|\begin{array}{c} (\lambda,\mu)_B \\ \alpha~L~L\end{array}\right\rangle\equiv
|\beta n_0 n_1 n_2 \alpha \rangle
=c_1^\beta c_2^{n_0} c_3^{n_1}c_4^{n_2}c_5^\alpha|0 \rangle,\label{bm1n}
\end{eqnarray}
where
$c_1=w$, $c_2=\xi_1$, $c_3=x_{10}$, $c_4=\bxi^2$, $c_5=A$, and the powers
$n_0=L-\mu+2\alpha$,
$n_1=\mu-2\alpha-\beta$,
$n_2=(\lambda+\mu-L-2\alpha-\beta)/2$,
and $\beta$ are defined by relations (\ref{bm1a}) and (\ref{betadef}).
The expression $\xi_{-1},\xi_1,\eta_{-1},\eta_0$ in terms of $c_1,...,c_5$ give
us the formal expressions: $\xi_1=c_2$, $\eta_0=c_3/c_2+\xi_0 \eta_1/c_2$,
$\xi_{-1}=-c_4/2/c_2+\xi_0^2/2/c_2$, $\eta_{-1}=(c_1-(-\xi_0 c_3-(1/2) \xi_0^2
\eta_1+(1/2) \eta_1 c_4))/c_2^2$, and the following relation among the
components $c_1,...,c_5$:
\begin{eqnarray*}
c_1^2=-c_5c_2^2+c_4c_3^2.
\end{eqnarray*}
Firstly, we calculate action of operators $Q_{0}$, $L_{-1}$ and $L_{-1}^2$ over
eigenvectors (\ref{bm1n}) and express them in terms of components $c_1,...,c_5$.
Using the above relations, implemented in REDUCE we obtain action of operators
$Q_{0}$, $L_{-1}$ and $L_{-1}^2$ onto eigenvectors (\ref{bm1n}). For even
$\lambda{+}\mu{-}L$ we get:
\begin{eqnarray*}
&&Q_0\left|\begin{array}{c} (\lambda,\mu)_B \\ \alpha~L~L\end{array}\right\rangle
\nonumber=[6\alpha c_2^{n_0-2}c_3^{n_1+2}c_4^{n_2+1}c_5^{\alpha -1}
+6\alpha \xi_0^2c_2^{n_0-2}c_3^{n_1+2}c_4^{n_2}c_5^{\alpha -1}
+6n_2\xi_0^2c_2^{n_0}c_3^{n_1}c_4^{n_2-1}c_5^{\alpha }
\\\nonumber&&\quad+(-4\alpha -n_0+n_1-2n_2)c_2^{n_0}c_3^{n_1}c_4^{n_2}c_5^{\alpha }
+12\alpha \xi_0c_1c_2^{n_0-2}c_3^{n_1+1}c_4^{n_2}c_5^{\alpha -1}
]|0\rangle
\\ %\end{eqnarray*}
%\begin{eqnarray*}
&&L_{{-}1}\left|\begin{array}{c} (\lambda,\mu)_B \\ \alpha{{+}}s~L{{+}}1~L{{+}}1\end{array}\right\rangle
\nonumber =[(-n_1+1) c_2^{n_0+2}c_3^{n_1-2}c_4^{n_2-1}c_5^{\alpha +1}
+n_1   c_2^{n_0}c_3^{n_1}c_4^{n_2}c_5^{\alpha }
\\\nonumber&&\quad+(n_0+n_1+1)\xi_0c_1c_2^{n_0}c_3^{n_1-1}c_4^{n_2-1}c_5^{\alpha }]|0\rangle
%\end{eqnarray*}
%\begin{eqnarray*}
\\
&& (L_{{-}1})^2\left|\begin{array}{c} (\lambda,\mu)_B \\ \alpha{{+}}s~L{{+}}2~L{{+}}2\end{array}\right\rangle
 \nonumber =[(2n_0^2+4n_0n_1+7n_0+2n_1^2+7n_1+6)/2\xi_0^2c_2^{n_0}c_3^{n_1}c_4^{n_2-1}c_5^{\alpha
}
\\\nonumber&&\quad+(-n_1^2+n_1)c_2^{n_0+2}c_3^{n_1-2}c_4^{n_2-1}c_5^{\alpha +1}
+(-n_0+2n_1^2-n_1-2)/2c_2^{n_0}c_3^{n_1}c_4^{n_2}c_5^{\alpha }
\\\nonumber&&\quad+(2n_0n_1+2n_1^2+3n_1)\xi_0c_1c_2^{n_0}c_3^{n_1-1}c_4^{n_2-1}c_5^{\alpha
}]|0\rangle
\end{eqnarray*}
and for odd $\lambda{+}\mu{-}L$ we get:
\begin{eqnarray*}
 &&Q_0\left|\begin{array}{c} (\lambda,\mu)_B \\ \alpha~L~L\end{array}\right\rangle
\nonumber = [12\alpha \xi_0c_2^{n_0-2}c_3^{n_1+3}c_4^{n_2+1}c_5^{\alpha-1}
+(-12\alpha -6)\xi_0c_2^{n_0}c_3^{n_1+1}c_4^{n_2}c_5^{\alpha }
\\\nonumber&&\quad+6\alpha c_1c_2^{n_0-2}c_3^{n_1+2}c_4^{n_2+1}c_5^{\alpha -1}
+6\alpha \xi_0^2c_1c_2^{n_0-2}c_3^{n_1+2}c_4^{n_2}c_5^{\alpha -1}
+6n_2\xi_0^2c_1c_2^{n_0}c_3^{n_1}c_4^{n_2-1}c_5^{\alpha }
\\\nonumber&&\quad+(-4\alpha -n_0+n_1-2n_2-3)c_1c_2^{n_0}c_3^{n_1}c_4^{n_2}c_5^{\alpha }]|0\rangle
%\end{eqnarray*}
%\begin{eqnarray*}
%&&
\\ && L_{{-}1}\left|\begin{array}{c} (\lambda,\mu)_B \\ \alpha{{+}}s~L{{+}}1~L{{+}}1\end{array}\right\rangle
\nonumber =[(n_0+n_1+2)\xi_0  c_2^{n_0}c_3^{n_1+1}c_4^{n_2}c_5^{\alpha }
+(n_1+1)   c_1c_2^{n_0}c_3^{n_1}c_4^{n_2}c_5^{\alpha }]|0\rangle
%\end{eqnarray*}
%\begin{eqnarray*}
%&&
\\
 && (L_{{-}1})^2\left|\begin{array}{c} (\lambda,\mu)_B \\ \alpha{{+}}s~L{{+}}2~L{{+}}2\end{array}\right\rangle
 \nonumber =[(-2n_0n_1-2n_1^2-5n_1)\xi_0c_2^{n_0+2}c_3^{n_1-1}c_4^{n_2-1}c_5^{\alpha
+1}
\\\nonumber&&\quad+(2n_0n_1+2n_0+2n_1^2+7n_1+5)\xi_0c_2^{n_0}c_3^{n_1+1}c_4^{n_2}c_5^{\alpha
}
+(-n_1^2+n_1)c_1c_2^{n_0+2}c_3^{n_1-2}c_4^{n_2-1}c_5^{\alpha +1}
\\\nonumber&&\quad+(2n_0^2+4n_0n_1+11n_0+2n_1^2+11n_1+15)/2\xi_0^2c_1c_2^{n_0}c_3^{n_1}c_4^{n_2-1}c_5^{\alpha
}
\\\nonumber&&\quad+(-n_0+2n_1^2+3n_1-1)/2c_1c_2^{n_0}c_3^{n_1}c_4^{n_2}c_5^{\alpha }]|0\rangle.
\end{eqnarray*}

Using the above actions of the operators $Q_{0}$, $L_{-1}$ and $L_{-1}^2$ on
eigenvectors (\ref{bm1n}) and extracting the coefficients at $c_1,...,c_5$, we
arrive to a set of equations with respect to unknown coefficients:
$\tilde a_{1}^{(0)}$, $\tilde a_{0}^{(0)}$, $\tilde a_{{-}1}^{(0)}$,
$\tilde a_{0}^{(1)}$, $\tilde a_{1}^{(1)}$, $\tilde a_{{-}1}^{(1)}$,
$\tilde a_{0}^{(2)}$, $\tilde a_{1}^{(2)}$, $\tilde a_{1}^{(2)}$ \ .
From formula (\ref{bm2n}) and from action of the operators $Q_{0}$, $L_{-1}$ and
$L_{-1}^2$ on functions (\ref{bm1n}) we obtain:
\begin{eqnarray}&&
Q_0\left|\begin{array}{c} (\lambda,\mu)_B \\
\alpha~L~L\end{array}\right\rangle-\sum_{s=0,,\pm1}\left\{
\tilde a_s^{(0)}\left|\begin{array}{c} (\lambda,\mu)_B \\
\alpha{+}s~L~L\end{array}\right\rangle
%\right.\\ \left.
{+}\tilde a_s^{(1)}L_{-1}\left|\begin{array}{c} (\lambda,\mu)_B \\
\alpha{+}s~L{+}1~L{+}1\end{array}\right\rangle
\nonumber\right.\\&& \qquad\qquad\qquad \left.
{+}\tilde a_s^{(2)}(L_{-1})^2\left|\begin{array}{c} (\lambda,\mu)_B \\
\alpha{+}s~L{+}2~L{+}2\end{array}\right\rangle
\right\}=0.\label{bm6n}
\end{eqnarray}
The solution of the set of equations obtained by extraction of coefficients
at the same powers of the operators $c_1$,...,$c_5$ gives values of unknown
coefficients
$\tilde a_{1}^{(0)}$, $\tilde a_{0}^{(0)}$, $\tilde
a_{{-}1}^{(0)}$, $\tilde a_{0}^{(1)}$, $\tilde a_{1}^{(1)}$, $\tilde
a_{{-}1}^{(1)}$, $\tilde a_{0}^{(2)}$, $\tilde a_{1}^{(2)}$, $\tilde
a_{1}^{(2)}$.
Finally, from (\ref{bm7n}) we arrive to needed values of
$ a_{1}^{(0)}$, $ a_{0}^{(0)}$, $ a_{{-}1}^{(0)}$, $ a_{0}^{(1)}$,
$ a_{1}^{(1)}$,
$ a_{{-}1}^{(1)}$, $ a_{0}^{(2)}$, $ a_{1}^{(2)}$, $ a_{1}^{(2)}$.
In result, the action of zero component $Q_0$ of the quadrupole operator onto the nonorthogonal
BM basis (\ref{bm1}), i.e. the coefficients $a_{s}^{(k)}$ of expansion (\ref{Q0-ask}), reads
as:
\begin{eqnarray}
\label{A2.4}
&&a^{(2)}_{0}{=}\frac{6(\lambda+\mu-L-2\alpha-\beta)}{((L+2)(2L+3))^{1/2}},\,\,\,
a^{(2)}_{-1}{=}\frac{12\alpha}{((L+2)(2L+3))^{1/2}},\,\,\,a^{(2)}_{-1}{=}0,
\\
&&
a^{(1)}_{0}{=}-6\frac{2\alpha\beta(L+2\alpha-\mu+1)+(\lambda+\mu-L-2\alpha)
{[\mu-2\alpha]}}
{(L+2)(L+1)^{1/2}}-\frac{6\beta}{(L+1)^{1/2}},\,
\nonumber\\
&&
a^{(1)}_{-1}{=}\frac{12\alpha(
{[L]}-\mu+2\alpha)}{(L+2)(L+1)^{1/2}},\,\,\,
a^{(1)}_{1}{=}\frac{6\beta(\lambda+\mu-L-2\alpha-\beta)
{[\mu-2\alpha-\beta]}}
{(L+2)(L+1)^{1/2}} ,
\nonumber\\
&& a^{(0)}_{0}{=}4\alpha\frac{L(L+1){-}3(L+2\alpha-\mu+\beta)^{2}}{(L+1)(2L+3)}
{-}2(\lambda+\mu-L-\beta-2\alpha)\frac{L(L+1){-}3(\mu{-}2\alpha)^{2}}{(L+1)(2L+3)}
\nonumber\\
&&
-(L-[2\mu]+4\alpha+\beta)\left( 1+\frac{3\beta}{L+1}\right),
\nonumber \\
&&
a^{(0)}_{-1}{=}\frac{
{[12\alpha]}(L-\mu+2\alpha)(L-\mu+2\alpha-1)}{(L+1)(2L+3)},
\nonumber \\
&& a^{(0)}_{1}{=}-\frac{6(\lambda+\mu-L-2\alpha-\beta)(\mu-2\alpha-\beta)(\mu-2\alpha-\beta-1)}{(L+1)(2L+3)},
\nonumber \\
&& \beta=\left\{
\begin{array}{l}
	0, \quad \lambda + \mu -L \mbox{ even},\\
	1, \quad \lambda + \mu -L \mbox{ odd}.
\end{array}
\right.
\nonumber
\end{eqnarray}
Note, there are five misprints in formulas (2.4) of Ref. \cite{Raychev1981},
that are corrected in the above expressions (\ref{A2.4}). They are marked by the
square brackets.

{\it Output of Step 6.} The set of matrix elements
$q_{i,j,k}^{(\lambda,\mu)}(L)$ has been calculated with algorithm implemented in
Fortran. The obtained results for $\lambda,\mu =0,...,10$, where
$\lambda\geq\mu$ (the corresponding sets of values $\alpha$ and $L$ (for example
see tables 1 of ref. \cite{casc18})) coincide up to 10 digits with results of
calculations obtained with help of symbolic algorithm \cite{casc18} implemented
in Wolfram Mathematica.  Note that the coefficients
$q_{i,j,k}^{(\lambda,\mu)}(L)$ for up to $\mu=3$ were calculated as well and
their values are equal to those presented in Table 1 of Ref. \cite{Raychev1981}
except values for the following sets of indices: $k=0$,$\mu=3$:
$q_{1,1,k}^{(\lambda,\mu)}(L)$ for $\lambda-L$ even,
$q_{1,0,k}^{(\lambda,\mu)}(L)$ and $q_{0,1,k }^{(\lambda,\mu)}(L)$ for
$\lambda-L$ odd; $k=1$, $\mu=3$: $q_{1,1,k}^{(\lambda,\mu)}(L)$ for $\lambda-L$
odd, $q_{0,0,k}^{(\lambda,\mu)}(L)$ and $q_{1,0,k}^{(\lambda,\mu)}(L)$ for
$L=\lambda+1$, $\lambda-L$ odd, $q_{0,0,k}^{(\lambda,\mu)}(L)$ for
$L=\lambda+2$, $\lambda-L$ even.  The corrected expressions are shown in
Appendix.

%%%%%%%%%%%%%%%%%%%%%%%%%%%%%%%%%%%%%%%%%%%%%%%%%%%%%%%%%%%%%%%%%%%%%%%%%
\subsection{ Calculations of action of the quadrupole operator onto      the
  orthonormalized BM basis}

The matrix elements of the quadrupole operators, generators of the group
$\Group{SU(3)}$, can be expressed as the reduced matrix elements by means of the
Wigner--Eckart theorem
\begin{equation}
\label{w-E}
\MatElemSU{(\lambda,\mu)\\j,L+k,M}{Q_m}{(\lambda,\mu)\\i,L,M'}=
\frac{(LM'\, 2m|L+k,M)}{\sqrt{2(L+k)+1}}
\RedMatElemSU{(\lambda,\mu)\\j,L+k}{Q}{(\lambda,\mu)\\i,L}.
\end{equation}
The corresponding reduced matrix elements are determined by the formula
 \begin{equation} \label{MatrixElemQ}
\RedMatElemSU{(\lambda,\mu)\\j, L+k}{Q}{(\lambda,\mu)\\i,L}
 =(-1)^k \frac{\sqrt{2L+1}}{(L+k,L,20|LL)}
q^{(\lambda,\mu)}_{i,j,k}(L),
 \end{equation}
 where the coefficients $q^{(\lambda,\mu)}_{i,j,k}(L)$ are defined by
 (\ref{qijk}).  In this definition, $k\geq 0$. Dimension of subspace of { the}
 ket vectors $|(\lambda,\mu) iLM\rangle$ at fixed $\lambda$ and $\mu$ are defined
 by formula (\ref{dimfa}).  {The dimension of this subspace determines the
   complexity of the above algorithms, i.e., requirements on the computer memory
   and execution time}.

\section{Conclusions} We present the practical
 algorithm implemented in
 Fortran for constructing the non-canonical {Bargmann--Moshinsky} (BM) basis with the highest weight vectors of
$\mathrm{SO(3)}$ irreps.,
 which can be used for calculating spectra and electromagnetic
transitions in molecular and nuclear physics.
The orthonormalisation algorithm \cite{Gantmacher} applied in Section 3, as well as recursion algorithm \cite{casc18}, allows one to calculate the orthonormalized BM basis in Fortran and Wolfram Mathematica, respectively.  The distinct advantage of
such orthonormalisation is that it does not involve any square root operation on the
expressions coming from the previous recursion steps of the conventional Gram--Schmidt algorithm.
This makes the proposed method very suitable for large-scale calculations
of spectral characteristics (especially close to resonances) of
quantum systems under consideration and to study their analytical properties for
understanding the dominant symmetries.
   {The formalism of partner groups $G=\mathrm{SU(3)} \times \overline{\mathrm{SU(3)}}$ allows for simulation of the intrinsic
properties of quantum systems (also nuclei), including their intrinsic symmetries.
The  presented nuclear SU(3) model is
extended and allows for additional intrinsic structure, especially it allows to
construct terms having required point symmetries.}
  Calculations of spectral characteristics
of the above nuclei models and study of their dominant symmetries will
be done in our next publications.

 The work was partially supported by the Bogoliubov-Infeld program, Votruba-Blokhintsev program, the RUDN University Program 5-100
and grant of Plenipotentiary of the Republic of Kazakhstan in JINR. AD\&PMK are grateful to Prof. A. G\'o\'zd\'z for hospitality during  visits in Institute  of Physics, Maria Curie{-}Sk{\l}odowska University.
\section*{Appendix}
  There are misprints in expressions of the coefficients $q_{i,j,k}^{(\lambda,3)}$ in Table 1 of Ref. \cite{Raychev1981},
that are corrected in the below expressions.  For  $\lambda-L$ even:
$$
q^{(\lambda,3)}_{1,1,0}{=}\left( \frac{12({-}2{+}L)({-}1{+}L)(2{-}L{-}\lambda)(6{+}L{+}\lambda)}{Y_{3}(\lambda,L)}
{+}6(2{+}\lambda){-}L(3{+}2\lambda{+}L(15{+}2\lambda))\right)
\frac{1}{(1{+}L)(3{+}2L)}.
$$
For  $\lambda{-}L$ odd:
$$
q^{(\lambda,3)}_{1,0,0}{=}q^{(\lambda,3)}_{0,1,0}{=}{-}
\frac{12\left(3(2{+}\lambda)(3{+}\lambda)(4{+}L{+}\lambda)(3{-}L{+}\lambda)({-}2{+}L)({-}1{+}L)(3{+}L)(2{+}L)\right)^{1/2}
}{\tilde {Y}_{3}(\lambda,L)(1{+}L)(3{+}2L)},
$$
$$
q^{(\lambda,3)}_{1,1,1}{=}
\frac{6(2({-}1{+}L)(7{+}L{+}\lambda){-}Y_{3}(\lambda,L{+}1))}{(1{+}L)}
\left(\frac{L(2{+}\lambda)(1{-}L{+}\lambda)}{\tilde{Y}_{3}(\lambda,L){Y}_{3}(\lambda,L{+}1)
(2{+}L)(3{+}2L)}\right)^{1/2}.
$$
For $L=\lambda{+}1$ and $\lambda{-}L$ odd:
$$
q^{(\lambda,3)}_{0,0,1}{=}\frac{30(4{+}\lambda)}{(2{+}\lambda)}\left(\frac{6({-}1{+}\lambda)}
{(3{+}\lambda)(20{+}\lambda)(5{+}2\lambda)}\right)^{1/2},
$$
$$
q^{(\lambda,3)}_{1,0,1}{=}12\left(\frac{\lambda(4{+}\lambda)}
{\tilde{Y}_{3}(\lambda,\lambda{+}1)(2{+}\lambda)}\right)^{1/2}.
$$
For $L=\lambda{+}2$ and $\lambda{-}L$ even:
$$
q^{(\lambda,3)}_{0,0,1}{=}{-}\frac{6(3\lambda)^{1/2}}{(3{+}\lambda)},
$$
where
$$Y_3(\lambda,L){=}4(\lambda{+}L{+}6)(\lambda{-}L{+}2){+}3(L{+}2)(L{+}3),$$
$$\tilde Y_3(\lambda,L){=}4(\lambda{+}L{+}5)(\lambda{-}L{+}3){+}(L{+}2)(L{+}3).$$
Thus, table 1 of Ref. \cite{Raychev1981} together with the above expressions have been used
to test the {\it Step 6} of the proposed procedure implemented in Fortran.


\begin{thebibliography}{9}
\bibitem{Cseh15}
Cseh J 2015
%\textit{Algebraic models for shell-like quarteting of nucleons}
\textit{Phys. Lett. B} \textbf{743} 213
%
\bibitem{Dytrych16}
 Dytrych T, et al 2016
 %\textit{Efficacy of the SU(3) scheme for ab initio large-scale calculations beyond the lightest nuclei}
  \textit{Comput. Phys. Commun. }\textbf{207} 202
%
\bibitem{Ell58}
Elliott J P 1958
%\textit{Collective Motion in the Nuclear Shell Model I}
 \textit{Proc. R. Soc. Lond. A} \textbf{245} 128
\bibitem{Gozdz18} G\'o\'zd\'z A ,  P\c{e}drak A, Gusev A A and Vinitsky S I 2018
%\textit{Point symmetries in the nuclear SU(3) partner groups model}
\textit{Acta Phys. Polonica B Proc. Suppl.} \textbf{11} 19
\bibitem{Harvey}  Harvey, M 1968 \textit{The Nuclear SU$_3$ Model}
 Advances in Nuclear Physics Springer Eds: Baranger M and Vogt E (Boston, MA)
\bibitem{BarMos61}
 Bargmann V and  Moshinsky M 1961
 %\textit{Group theory of harmonic oscillators (II)}
\textit{ Nucl. Phys.} \textbf{23} 177
\bibitem{MosPatShaWin75}
 Moshinsky M, Patera J,  Sharp R T and  Winternitz P 1975
 %\textit{Everything you always wanted to know about $SU(3)\supset O(3)$}
\textit{ Ann Phys (NY)} \textbf{95} 139
\bibitem{Varshalovitch}
Varshalovitch D A, Moskalev A N and Hersonsky V K 1988
\textit{Quantum Theory of Angular Momentum}
 (Singapore: World Sci.)
\bibitem{Dudek2002}
Dudek J, Go\'zd\'z  A, Schunck N and  Mi\'skiewicz M 2002
%\textit{Nuclear Tetrahedral Symmetry: Possibly Present Throughout the Periodic Table}
\textit{Phys. Rev. Lett.} \textbf{88}   252502
\bibitem{Pan16}
 Pan  F,  Yuan S,   Launey K D and Draayer J P 2016
 %\textit{A new procedure for constructing basis vectors of $SU(3)\supset SO(3)$}
\textit{  Nucl. Phys A} \textbf{743} 70
\bibitem{AliRayRos81}
 Alisauskas S,  Raychev P and Roussev R 1981
%\textit{Analytical form of the orthonormal basis of the decomposition $SU(3)\supset O(3)\supset
%    O(2)$ for some $(\lambda,\mu)$ multiplets}
    \textit{J. Phys. G } \textbf{7}  1213
\bibitem{casc18} Deveikis A, Gusev A A, Gerdt V P, Vinitsky S I, G\'o\'zd\'z A and P\c edrak A 2018
 %Symbolic Algorithm for Generating the Orthonormal Bargmann--Moshinsky Basis for SU(3)Group
    %In: Gerdt V, Koepf W, Seiler W, Vorozhtsov E (eds) Computer Algebra in Scientific Computing CASC 2018
    \textit{Lect. Notes  Computer Sci.}   \textbf{11077} 131
%\bibitem{iopartnum} IOP Publishing is to grateful Mark A Caprio, Center for Theoretical Physics, Yale University, for permission to include the {\tt iopart-num} \BibTeX package (version 2.0, December 21, 2006) with  this documentation. Updates and new releases of {\tt iopart-num} can be found on \verb"www.ctan.org" (CTAN).
\bibitem{Afanasjev}
 Afanasjev  G N, Avramov S A and Raychev P P 1973
 %\textit{Realization of the Physical Basis for SU(3) and the Probabilities of E2 Transitions in the SU(3) Formalism}
\textit{Sov. J. Nucl. Phys.} \textbf{16}  53
\bibitem{Raychev1981}  Raychev P and Roussev  R 1981
%Matrix elements of the generators of SU(3) and of the basic O(3)
%scalars in the enveloping algebra of SU(3)
\textit{J. Phys. G} \textbf{7} 1227
\bibitem{Gantmacher} Gantmacher F R 1984  \textit{The Theory of Matrices} Vol. 1 (New York: Chelsea)

\end{thebibliography}
\end{document}